\documentclass[11pt]{article}
\usepackage{amsmath,amssymb}
\usepackage{amsfonts}
\usepackage{eucal}
\title{\leavevmode\vadjust{\vskip -5mm}
\textbf{The phase symmetry  \\
of classical electrodynamics}}
\author{\sc Serhii Samokhvalov\thanks{{\bf e}-{\it mail}: serh.samokhval@gmail.com}\\
 \tabaddress{Dniprovsk State Technical University, Ukraine}}

\date{ ***\\
} 

\def\tabaddress#1{{\small\it\begin{tabular}[t]{c}#1
\\[1.2ex]\end{tabular}}}

\def\<#1>{\langle#1\rangle}

\def\beq{\begin{equation}}
\def\eeq{\end{equation}}
\def\bea{\begin{eqnarray}}
\def\eea{\end{eqnarray}}
\def\beann{\begin{eqnarray*}}
\def\eeann{\end{eqnarray*}}
\def\ben{\begin{enumerate}}
\def\een{\end{enumerate}}

\def\qed{\ifvmode\removelastskip\fi
{\unskip\nobreak\hfil\penalty50\hbox{}\nobreak\hfil \hbox{\vrule
height1.2ex width1.2ex}\parfillskip=0pt \finalhyphendemerits=0
\par\smallskip}}

\def\texthook{\vrule height 0pt depth 0.4pt width 3.5pt
          \vrule height 5pt depth 0.4pt \kern 3pt}
\def\scripthook{\vrule height 0pt depth 0.2pt width 1.5pt
                \vrule height 3pt depth 0.2pt width 0.2pt \kern 1pt}

\begin{document}

\maketitle

\thispagestyle{empty}

\begin{abstract}

The dynamic $U (1)$ symmetry of classical electrodynamics without charges, similar to the known phase symmetry of quantum mechanics, is analyzed. Using this symmetry, alternative Lagrangians of classical electrodynamics, similar to quantum-mechanical ones, were found, and their symmetry properties were investigated.

\end{abstract}

\clearpage


\section{Introduction}

The most characteristic feature of quantum mechanics is the use of complex numbers to describe states, which leads to the existence of a compact phase of the wave function. The wave property of matter are connected with it. A notable peculiarity of the quantum-mechanical phase usage is the existence of the phase symmetry (the requirement of the unitarity of the quantum theory).

Besides the wave characteristics, the quantization of the particles quantity is the result of the phase compactness and their conservation is connected with the phase symmetry.

Thus, the two main properties of quantum mechanics: wave-likeness and discreteness there are due to phase symmetry, that requires the comprehension of the role of phase symmetry in it to understand quantum-mechanical laws.

We will look for such comprehension by comparing quantum mechanics with classical electrodynamics, which, unlike quantum mechanics, was formulated from the very beginning as purely real, while at the same time having deep analogies with quantum mechanics.

Greek indices take on values $0,1,2,3$, Latin indexes of the middle of the alphabet take on values $1,2,3$, $diag(g_{\mu\nu})=(-,+,+,+)$, for a completely anisymmetric tensor is accepted $\varepsilon^{0123}=1$.

\vskip 3mm

\section{The phase in the quantum mechanics}

The wave function meets Schr\"{o}dinger equation:

\begin{equation}\label{eq1}
i\hbar\partial_t \psi =-\frac{\hbar^2}{2m} \triangle\psi +V\psi.
\end{equation}

If represent

\begin{equation}\label{eq2}
\psi=R \exp \left(i\theta /\hbar
\right)=\rho^{\frac{1}{2}}\left(\cos\frac{\theta}{\hbar}+i
\sin\frac{\theta}{\hbar}\right),
\end{equation}

\noindent where $\rho=R^2 =\exp (2\tau)$, complex equation
(\ref{eq1}) splits into two real equations (its real and imagine parts), i.e. its decomplexification takes place:

\begin{equation}\label{eq3}
\underline{Re}: \qquad \partial_t \theta =-\frac{1}{2m}(\nabla\theta)^2
-V-Q,
\end{equation}

\begin{equation}\label{eq4}
\underline{Im}: \qquad\quad \; \partial_t \rho
+\nabla\left(\rho\frac{1}{m}\nabla\theta\right)=0,
\end{equation}
where
\begin{equation}\label{eq5}
Q =-\frac{\hbar^2}{2m}\frac{\triangle
R}{R}=\frac{\hbar^2}{4m}\left[\frac{1}{2}\left(\frac{\nabla\rho}{\rho}\right)^2-\frac{\nabla\rho}{\rho}\right]
\end{equation}

\noindent -- quantum-mechanical potential. That is it that makes a distinction between quantum-mechanical and classical descriptions, and it's caused by the superposition principle, i.e. by Schr\"{o}dinger equation linearity (\ref{eq1}). Equation (\ref{eq3}) is Hamilton-Jacobi equation, and equation (\ref{eq4}) is equation of particles transfer at velocity $\vec v=\nabla\theta/m$, i.e. $\vec p=\nabla\theta$.

From equation (\ref{eq3}) comes directly the equation of motion for a particle, if we take gradient from each of its parts

\begin{equation}\label{eq6}
\partial_t \vec p+\vec v\nabla\vec p =-\nabla U
\end{equation}

\noindent (in Lagrangian coordinates $d\vec p/dt=-\nabla U$) and denote $U=V+Q$. So a particle is affected by quantum-mechanical force $-\nabla Q$ in addition to classical force $-\nabla V$.

If instead of $\rho$ we use variable $\tau$ through which the quantum-mechanical potential is put down like

\begin{equation}\label{eq7}
Q=-\frac{\hbar^2}{2m}\left[(\nabla\tau)^2 +\nabla\tau\right],
\end{equation}

\noindent equations (\ref{eq3}) and (\ref{eq4}) will become:

\begin{equation}\label{eq8}
\partial_t \theta =-\frac{1}{2m}(\nabla\theta )^2 -V +\frac{\hbar^2}{2m}\left[(\nabla\tau)^2
+\nabla\tau\right],
\end{equation}

\begin{equation}\label{eq9}
\partial_t \tau +\frac{1}{m}\nabla\theta\nabla\tau
=-\frac{1}{2m}\nabla\theta.
\end{equation}

All the equations and definitions listed above are symmetrical in respect of phase transformations:

\begin{equation}\label{eq10}
\delta\theta=\hbar\delta\varphi.
\end{equation}

With direct (algebraic) decomplexification $\psi=\varepsilon
+i\chi$ Schr\"{o}dinger equation (\ref{eq1}) splits into two
equations:

\begin{equation}\label{eq11}
\underline{Re}:\qquad \hbar\partial_t \chi
=-\frac{\hbar^2}{2m}\triangle\varepsilon -V\varepsilon,
\end{equation}

\begin{equation}\label{eq12}
\underline{Im}:\qquad \hbar\partial_t \varepsilon
=-\frac{\hbar^2}{2m}\triangle\chi +V\chi.
\end{equation}

Phase symmetry for equations (\ref{eq11}) and (\ref{eq12}) is put
down as:

\begin{equation}\label{eq13}
\partial\varepsilon =-\chi\delta\varphi, \qquad \delta\chi =\varepsilon\delta\varphi.
\end{equation}\vskip 3mm

\section{Maxwell equations complexification}

Maxwell equations in space without charges are:

\begin{equation}\label{eq14}
\frac{1}{c}\partial_t \vec H =-\nabla \times \vec E, \qquad \nabla
\cdot \vec H =0,
\end{equation}

\begin{equation}\label{eq15}
\frac{1}{c}\partial_t \vec E =\nabla \times \vec H, \qquad \nabla
\cdot \vec E =0.
\end{equation}
Equations (\ref{eq14}) are the first pair and equations (\ref{eq15}) are the second pair of Maxwell equations. Equations(\ref{eq14}) are consequence of expression $\vec E$ and $\vec H$ through electromagnetic potentials (Jacobi identities).

Equations (\ref{eq14}) and (\ref{eq15}) are symmetric with respect to Rainich transformations \cite{1}, which form group $U(1)^R$:

\begin{equation}\label{eq16}
\delta\vec E =-\vec H\delta\varphi, \qquad \delta\vec H =\vec E
\delta\varphi,
\end{equation}

\noindent which are the analogue of phase transformations (\ref{eq13}) in quantum mechanics. This symmetry is completely dynamic and it doesn't exist in the presence of charges (though symmetry (\ref{eq13}) take place also with the presence of the potential $V$). Transformations (\ref{eq16}) confuses the first and the second pairs of Maxwell equations, i.e. they confuses Lagrangian equation for Maxwell's Lagrangian and Jacobi identity for $F_{\mu\nu}$.

This symmetry lets to put down Maxwell equations in complex form:

\begin{equation}\label{eq17}
\frac{i}{c}\partial_t \vec \xi =\nabla\times \vec \xi,
\end{equation}

\begin{equation}\label{eq18}
\nabla\cdot \vec \xi =0,
\end{equation}

\noindent where $\vec \xi =\vec E + i\vec H$. If equation
(\ref{eq17}) is written in components:

\begin{equation}\label{eq19}
\frac{i}{c}\partial_t \xi^i -\varepsilon^{ijk}\partial_j \xi_k =0,
\end{equation}

\noindent it look a lot like Dirac equation.

Let's also introduce $\vec \eta =\vec \xi^{*} =\vec E -i\vec H$. Then

\begin{equation}\label{eq20}
\vec \xi\cdot\vec \eta = E^2 +H^2,
\end{equation}

\begin{equation}\label{eq21}
i(\vec \xi \times \vec \eta )=2(\vec E \times \vec H).
\end{equation}

\noindent Vector $\vec \eta$ obeys the equations:

\begin{equation}\label{eq22}
\frac{i}{c}\partial_t\vec \eta =-\nabla\times\vec \eta,\qquad
\nabla\cdot\vec \eta =0.
\end{equation}

Rainich transformations (\ref{eq16}) for fields $\vec \xi$ and $\vec \eta$ have such form:

\begin{equation}\label{eq23}
\delta\vec \xi =i\vec \xi \delta\varphi,\qquad \delta\vec \eta
=-i\vec \eta \delta\varphi,
\end{equation}

\noindent or for finite transformations:

\begin{equation}\label{eq24}
\vec \xi ' =e^{i\varphi}\vec \xi ,\qquad \vec \eta '
=-e^{-i\varphi}\vec \eta .
\end{equation}

Multiplying (\ref{eq17}) by $\vec \eta$, and (\ref{eq22}) by $\vec \xi$, and adding them, we get:

\begin{equation}\label{eq25}
\partial_t (\vec \xi\cdot\vec \eta)+\nabla\cdot\left[ic(\vec \xi\times\vec \eta)\right] =
0.
\end{equation}

\noindent Energy density and Umov-Pointing vector

\begin{equation}\label{eq26}
\omega =\frac{1}{8\pi}(\vec \xi\cdot\vec \eta) =\frac{1}{8\pi}(E^2 +H^2),
\end{equation}

\begin{equation}\label{eq27}
\vec S =\frac{ic}{8\pi}(\vec \xi\times\vec \eta)
=\frac{c}{4\pi}(\vec E +\vec H)
\end{equation}

\noindent let equation (\ref{eq25}) be written in the form of the
equation:

\begin{equation}\label{eq28}
\partial_t\omega +\nabla\cdot\vec S =
0.
\end{equation}

For any substance with density $\rho$ under its transfer velocity we have to take vector $\vec v$ that belongs to the transfer equation:

\begin{equation}\label{eq29}
\partial_t \rho +\nabla\cdot(\rho\vec v) =
0.
\end{equation}

\noindent Thus, the speed of propagation of electromagnetic energy is determined as follows:

\begin{equation}\label{eq30}
\vec v =\vec S/\omega =2c\frac{\vec E\times\vec H}{E^2 +H^2}
\end{equation}

\noindent and we have $v=c$ only in the case of electromagnetic wave when $\vec E\perp\vec H$ and $E=H$. In other cases $v\neq c$. For
example, for electrostatics $\vec H=0$ and $\vec v=0$. This is effect of mass.

Let the constant electric field $\vec \varepsilon$ be superimposed upon an electromagnetic wave $\vec E$, $\vec H$ (such a solution is physical, i.e. it definitely obeys Maxwell equations). Then we get from (\ref{eq30}):

\begin{equation}\label{eq31}
\vec v =\vec S/\omega =2c\frac{\vec E\times\vec H+\vec
\varepsilon\times\vec H}{E^2 +H^2 + 2\vec\varepsilon\cdot\vec E
+\varepsilon^2}.
\end{equation}

\noindent Let's assume that $\varepsilon$ is small and $\vec x=\vec\varepsilon /E$. From (\ref{eq31}) with the accuracy to the first order on $x$ we get:

\begin{equation}\label{eq32}
\vec v =c\left[(1-\vec x\cdot\vec n_E )\vec n_S +\vec x\times\vec
n_H\right],
\end{equation}

\noindent where $\vec n_E$ and $\vec n_H$  are unit vectors along vectors $\vec E$, $\vec H$ and $\vec n_S =\vec n_E\times\vec n_H$. From (\ref{eq32}) comes that the velocity of the an electromagnetic wave spreading in the constant electric field fluctuates with the wave frequency not only in the value (the first term in (\ref{eq32})) but also in the direction (the last term).\vskip 3mm

\section{Phase in electrodynamics}

On the analogy with (\ref{eq2}) let's put:

\begin{equation}\label{eq33}
\vec \xi = J^{1/2}\left(\vec n_E \cos\frac{\theta}{\hbar} +i\vec
n_H \sin\frac{\theta}{\hbar}\right),
\end{equation}

 \noindent i.e.

\begin{equation}\label{eq34}
E = J^{1/2}\cos\frac{\theta}{\hbar},\qquad H =
J^{1/2}\sin\frac{\theta}{\hbar},
\end{equation}

\noindent $J=E^2 +H^2$. Phase $\theta$ corresponds to the action for photons and at Rainich transformations (\ref{eq23}) obtains the addition:

\begin{equation}\label{eq35}
\delta\theta =\hbar \delta\varphi,
\end{equation}

\noindent which completely coincides with (\ref{eq10}).

Now our goal is in phase decomplexification of dynamic (with derivativs with respect to $t$) Maxwell equations (\ref{eq17}) by analogy with decomplexification of Schr\"{o}dinger equations. As a result we get:

\[
\underline{Re}: \qquad -\partial_t J \sin\frac{\theta}{\hbar}\vec
n_H -2\frac{J}{\hbar}\cos\frac{\theta}{\hbar}\partial_t \theta\vec
n_H -2J\sin\frac{\theta}{\hbar}\partial_t\vec n_H =\]

\begin{equation}\label{eq36}
 c \left\{
\cos\frac{\theta}{\hbar}(\nabla J\times\vec n_E)
-2\frac{J}{\hbar}\sin\frac{\theta}{\hbar}(\nabla\theta\times\vec
n_E) +2J\cos\frac{\theta}{\hbar}(\nabla\times\vec n_E)\right\},
\end{equation}

\[
\underline{Im}: \qquad \partial_t J \cos\frac{\theta}{\hbar}\vec
n_E -2\frac{J}{\hbar}\sin\frac{\theta}{\hbar}\partial_t \theta\vec
n_E +2J\cos\frac{\theta}{\hbar}\partial_t\vec n_E =\]

\begin{equation}\label{eq37}
 c \left\{
\sin\frac{\theta}{\hbar}(\nabla J\times\vec n_H)
+2\frac{J}{\hbar}\cos\frac{\theta}{\hbar}(\nabla\theta\times\vec
n_H) +2J\sin\frac{\theta}{\hbar}(\nabla\times\vec n_H)\right\}.
\end{equation}

Let's multiply (\ref{eq36}) by $\cos\frac{\theta}{\hbar}\vec n_H$, and (\ref{eq37}) by $\sin\frac{\theta}{\hbar}\vec n_E$ and add. As a result we get the analogue of Hamilton-Jacobi equation (\ref{eq3}) for photons:

\begin{equation}\label{eq38}
\partial_t\theta +\frac{\hbar}{2J}\nabla\cdot\left(J c \cos
\left(2\frac{\theta}{\hbar}\right)\vec
n_S\right)+\hbar\frac{c}{2}\left[\vec n_H \cdot (\nabla\times\vec
n_E )+\vec n_E\cdot (\nabla \times\vec n_H )\right]=0.
\end{equation}

\noindent If multiply (\ref{eq36}) by $-\sin\frac{\theta}{\hbar}\vec n_H$, and (\ref{eq37}) by $\cos\frac{\theta}{\hbar}\vec n_E$ and add, we'll get photons transfer equation:

\begin{equation}\label{eq39}
\partial_t J+\nabla\cdot\left(J c
\sin\left(2\frac{\theta}{\hbar}\right)\vec n_S \right) =0,
\end{equation}

\noindent which is the same as (\ref{eq28}). Equations (\ref{eq38}) and (\ref{eq39}) do not exhaust equations (\ref{eq36}) and (\ref{eq37}). Other ratios can be obtained from them. \vskip 3mm

\section{Lagrangians of electrodynamics}

Let's further suppose $c=1$, and also instead of index $t$ let's write index $0$.

Maxwell's Lagrangian of electromagnetic field

\begin{equation}\label{eq40}
L^M =-\frac{1}{16\pi}F_{\mu\nu}F^{\mu\nu}=\frac{1}{8\pi}(E^2 -H^2
) =\frac{1}{8\pi}Re\xi^2
\end{equation}

\noindent leads to second pair of Maxwell equations (\ref{eq15})

\begin{equation}\label{eq41}
\partial_\nu F^{\mu\nu}=0
\end{equation}

\noindent only provided we express $F_{\mu\nu}$ through potentials $A_\mu$:

\begin{equation}\label{eq42}
F_{\mu\nu}=\partial_\mu A_\nu -\partial_\nu A_\mu ,
\end{equation}

\noindent which we should vary. The first pair of Maxwell equations (\ref{eq14}):

\begin{equation}\label{eq43}
\partial_\nu \tilde F^{\mu\nu}=0,
\end{equation}

\noindent where

\begin{equation}\label{eq44}
\tilde F^{\mu\nu}=\frac{1}{2}\varepsilon^{\mu\nu\rho\tau}F_{\rho\tau},
\end{equation}

\noindent is Jacobi identity - the consequence of (\ref{eq42}).

In spite of $U(1)^R$-invariance of Maxwell equations, Lagrangian $L^M$ is not $U(1)^R$-invariant, though obviously Lorentz-invariant. Transformations (\ref{eq16}) really lead to the addition:

\begin{equation}\label{eq45}
\delta L^M =-\frac{1}{2\pi}\vec E\cdot \vec H\delta\varphi.
\end{equation}

Let us try to find the $ U (1) ^ R $-invariant Lagrangian of electrodynamics without worrying about its Lorentz invariance. At that let's use the analogy of dynamic Maxwell and Dirac equations, and write down Lagrangian of electrodynamics, which similar to Dirac's Lagrangian:

\begin{equation}\label{eq46}
L^D =-\frac{l}{8\pi}\left[\vec H\cdot (\partial_0 \vec E
-\nabla\times\vec H)-\vec E\cdot(\partial_0\vec H+\nabla\times\vec
E)\right].
\end{equation}
where $l$ is the constant of the length dimension.

Variational derivatives of $L^D$ with respect to $\vec E$ and $\vec H$ give the dynamic part of Maxwell equations both the first (\ref{eq14}) and the second (\ref{eq15}) pair:

\begin{equation}\label{eq47}
\frac{\delta L^D}{\delta\vec E}=\frac{l}{2\pi}(\partial_0 \vec
H+\nabla\times \vec E)=0,
\end{equation}

\begin{equation}\label{eq48}
\frac{\delta L^D}{\delta\vec H}=\frac{l}{2\pi}(\partial_0 \vec
E-\nabla\times \vec H)=0.
\end{equation}

\noindent Lagrangian $L^D$ don't give zero divergences for $\vec E$ and $\vec H$ as Lagrange equations.

Lagrangian $L^D$ is invariant with respect to transformations (\ref{eq16}) ($U(1)^R$ - invariant), though obviously is not invariant with respect to Lorenz transformations.

$U(1)^R$-invariance allows us to write down $L^D$ in the complex form (with respect to $\vec \xi$ and $\vec \eta$):

\begin{equation}\label{eq49}
L^D =-\frac{l}{16\pi}\left[\vec \eta\cdot (i\partial_0 \vec \xi
-\nabla\times\vec \xi)-\vec \xi\cdot(i\partial_0\vec
\eta+\nabla\times\vec \eta)\right].
\end{equation}

Independent variation on $\vec\xi$ and $\vec\eta$ obviously leads to equation (\ref{eq17}) and the first of the equations (\ref{eq22}). $U(1)^R$ - invariance in notation (\ref{eq49}) becomes utterly evident.\vskip 3mm

\section{Relativistic transformations of $L^D$}

At the Lorenz transformations with parameters $\vartheta^i$ (angles) and $v^i$ (velocities), we have:

\begin{equation}\label{eq50}
\delta E_i =\delta\vartheta_k\varepsilon_{kij}E_j -\delta v_k
\varepsilon_{kij}H_j ,
\end{equation}

\begin{equation}\label{eq51}
\delta H_i =\delta\vartheta_k\varepsilon_{kij}H_j -\delta v_k
\varepsilon_{kij}E_j ,
\end{equation}

\noindent or in the complex form:

\begin{equation}\label{eq52}
\delta \xi_i =(\delta\vartheta_k+i\delta v_k
)\varepsilon_{kij}\xi_j ,
\end{equation}

\begin{equation}\label{eq53}
\delta \eta_i =(\delta\vartheta_k+i\delta v_k
)\varepsilon_{kij}\eta_j .
\end{equation}

\noindent Beside this:

\begin{equation}\label{eq54}
\delta\partial_0=-\delta v_i \partial_i ,
\end{equation}

\begin{equation}\label{eq55}
\delta\partial_m=\delta v_m
\partial_0+\delta\vartheta_k\varepsilon_{kmi}\partial_i .
\end{equation}

\noindent Let's bring into use designations:

\begin{equation}\label{eq56}
R_s=i\partial_0\xi_s-\varepsilon_{smn}\partial_m\xi_n=\frac{4\pi}{l} \frac{\delta
L^D}{\delta\eta^s},
\end{equation}

\begin{equation}\label{eq57}
D=\partial_n\xi_n,
\end{equation}

\noindent relativeistic transformations of which are:

\begin{equation}\label{eq58}
\delta R_s=\delta\vartheta_k\varepsilon_{ksj}R_j-i\delta v_s D,
\end{equation}

\begin{equation}\label{eq59}
\delta D=-i\delta v_k R_k.
\end{equation}

\noindent With the use of $R_s$ we can write down:

\begin{equation}\label{eq60}
L^D=\frac{l}{4\pi}Re(\eta_s R_s),\qquad \delta L^D
=\frac{l}{4\pi}Re\left[\delta(\eta_sR_s)\right].
\end{equation}

\noindent But:

\begin{equation}\label{eq61}
\delta (\eta_sR_s)=i\delta v_m\eta_m(\varepsilon_{mnk}R_k-g_{mn}D)
\end{equation}

\noindent and when accomplishing Maxwell equations

\begin{equation}\label{eq62}
R_k=0,\qquad D =0,
\end{equation}

\noindent $\delta L^D$ disappears. It's interesting, that lorenzinvariance of electrodynamics, that is based on the Lagrangian $L^D$, demands $D=0$ as the extra condition, but $D=0$ isn't Lagrange equation for $L^D$.

Let's write $\delta L^D$ with help of $\vec E$ and $\vec H$:

\begin{equation}\label{eq63}
4\pi\delta L^D=\delta v_m \,l
\left[H_n(\varepsilon_{mnk}R^{Re}_{k}-g_{mn}D^{Re})-E_n(\varepsilon_{mnk}R^{Im}_{k}-g_{mn}D^{Im})\right],
\end{equation}

\noindent where

\begin{equation}\label{eq64}
R_s^{Re}=-\partial_0 H_s -\varepsilon_{smn}\partial_m E_n,\qquad
R^{Im}_s =\partial_0 E_s -\varepsilon_{smn} \partial_m H_n,
\end{equation}

\begin{equation}\label{eq65}
D^{Re}=\partial_n E_n,\qquad D^{Im} =\partial_n H_n.
\end{equation}

\noindent Substituting (\ref{eq64}) and (\ref{eq65}) in (\ref{eq63}), we get:

\[
8\pi\delta L^D=\delta v_m \,l\,
[-\varepsilon_{mnk}(H_n\partial_0 H_k+E_n\partial_0 E_k)+(H_k\partial_mE_k-E_k\partial_mH_k)+
\]

\begin{equation}\label{eq66}
 (E_k\partial_kH_m-H_k\partial_kE_m)+(E_m\partial_kH_k-H_m\partial_kE_k)].
\end{equation}

\noindent As we see, $\delta L^D$ doesn't transform into divergence on the level of the fields $\vec E$ and $\vec H$.

Both Lagrangians are invariant with respect to space-time translations \cite{2}:

\begin{equation}\label{eq67}
\delta x^\mu =t^\mu ,
\end{equation}

\begin{equation}\label{eq68}
\delta x^5 =A^\mu t^\mu ,
\end{equation}

\noindent ($x^5$ - is the fifth coordinate), that lead to the such transformations of the fields:

\begin{equation}\label{eq69}
\delta A_\mu =- t^\nu F_{\nu\mu},
\end{equation}

\begin{equation}\label{eq70}
\delta E_i =- t^\nu \partial_\nu E_{i},
\end{equation}

\begin{equation}\label{eq71}
\delta H_i =- t^\nu \partial_\nu H_{i}.
\end{equation}
\vskip 3mm

\section{Conserved currents}

Let's find currents whose conservation is ensured by the considered invariants of both Lagrangians.

If the Lagrangian is symmetric with respect to transformations:

\begin{equation}\label{eq72}
\delta x^\mu =\varepsilon^a X^{\mu}_a ,
\end{equation}

\begin{equation}\label{eq73}
\delta q^A =\varepsilon^a \alpha^{A}_a ,
\end{equation}

\noindent then, according to Noether's theorem, currents

\begin{equation}\label{eq74}
J^{\mu}_a=-\frac{\partial L}{\partial
(\partial_{\mu}q^A)}\alpha^A_a-LX^{\mu}_a.
\end{equation}
are conserved on the extremals.

For the Lagrangian $L^M$ in case of translations (\ref{eq67}) and (\ref{eq69}) we obtain:

\begin{equation}\label{eq75}
\alpha^A_a\sim -F_{\nu\mu},\qquad X^\mu_a\sim\delta^\mu_a
\end{equation}

\noindent and as a conserved current acts Maxwell's energy-momentum tensor

\begin{equation}\label{eq76}
T^\mu_\nu
=\frac{1}{4\pi}\left(-F^{\rho\mu}F_{\rho\nu}+\frac{1}{4}\delta^\mu_\nu
F_{\rho\tau}F^{\mu\tau}\right),
\end{equation}

\noindent for which  $T^0_0$ coincide with $\omega$ (\ref{eq26}),
and $T^i_0$ with $S^i$ (\ref{eq27}).

For the Lagrangian $L^D$ in case of translations (\ref{eq67}), (\ref{eq70}) and (\ref{eq71}), we obtain:

\begin{equation}\label{eq77}
\alpha^A_a\sim -\partial_\nu E_{i},\;\; -\partial_\nu H_i;\qquad
X^A_a\sim\delta^A_a.
\end{equation}
Then the density of the "energy" for the Lagrangian $L^D$ is:

\begin{equation}\label{eq78}
T^0_0 =\frac{l}{8\pi}\left\{\vec E\cdot[\nabla\times \vec E]+\vec
H \cdot[\nabla\times \vec H]\right\}.
\end{equation}

\noindent On shell:

\begin{equation}\label{eq79}
\nabla\times \vec E =-\partial_0\vec H,\qquad \nabla\times \vec H
=\partial_0\vec E
\end{equation}

\noindent and instead of (\ref{eq78}) we have:

\begin{equation}\label{eq80}
T^0_0 =-\frac{l}{8\pi}\left\{\vec E\cdot\partial_0 \vec H-\vec H
\cdot\partial_0 \vec E\right\},
\end{equation}

\noindent that practically coincides with density of energy for Dirac's field and, apparently, doesn't coincide with (\ref{eq26}). For example, for electrostatic field, (\ref{eq80}) gives $0$. Due to the fact that in the electromagnetic wave $\vec E\bot \vec H$, for the flat-polarized wave with (\ref{eq80}) we also get $0$. Very odd "energy"!

For the density of the "momentum" for the Lagrangian $L^D$ we have:

\begin{equation}\label{eq81}
T^i_0\vec e_i =\frac{l}{8\pi}\left\{[\vec E\times \partial_0\vec
E]+[\vec H\times \partial_0\vec H]\right\}.
\end{equation}

\noindent or on shell (\ref{eq79})

\begin{equation}\label{eq82}
T^i_0\vec e_i =\frac{l}{8\pi}\left\{[\vec E\times[\nabla\times
\vec H]]-[\vec H \times[\nabla\times \vec E]]\right\}.
\end{equation}

\noindent It's also very odd quantity. For the flat-polarized electromagnetic wave with (\ref{eq81}) we get $T^i_0=0$.

Now let's consider Rainich current, that is Noether current of the transformations (\ref{eq16}) for the Lagrangian $L^D$. In such case:

\begin{equation}\label{eq83}
\alpha^A_a\sim -H_i,\;\; E_i; \qquad X^\mu_a=0
\end{equation}

\noindent and current is:

\begin{equation}\label{eq84}
J^0=l\,\omega=\frac{l}{8\pi}(E^2+H^2),
\end{equation}

\begin{equation}\label{eq85}
J^i=l\,S^i=\frac{l}{4\pi}[\vec E\times \vec H]^i.
\end{equation}

It's interesting, that "phase" current for the Lagrangian $L^D$ coincides (up to the factor $l$) with "temporal" current for the Lagrangian $L^M$. What does it mean? Maybe, synchronous moving by phase happens when moving by time?

If quantity of photons is connected  with the $U(1)^R$-symmetry, then the formulas (\ref{eq84}) and (\ref{eq85}) give the "quantum-mechanical" interpretation of the light intensity. In such case the Lagrangian $L^D$ is better than $L^M$. The transition from $ L ^ M $ to $ L ^ D $ may correspond to "quantization".

Let's note, that when $\partial_0\vec E \sim -h\nu\vec H$ and $\partial_0\vec H \sim h\nu\vec E$ (which corresponds to the synchronous phase transformation during evolution over time), from (\ref{eq80}) and (\ref{eq81}) we obtain:

\begin{equation}\label{eq86}
T^0_0 \sim h\nu\omega,\qquad T^i_0\sim h\nu S^i.
\end{equation}

\noindent If $\omega$ is proportional to the quantity of the photons, and $S^i$ is proportional to the density of their flow, then obtained formulas lead to the quantum interpretation of the energy as a quantity that is proportional to $h\nu$, and of the momentum as a quantity that is proportional to $h/\lambda$.

Let's note, that Rainich phase transformations (\ref{eq16}) and translations along the fifth coordinate

\begin{equation}\label{eq87}
\delta x^5 =t^5 ,
\end{equation}

\noindent which are associated with the conservation of charges, in this formalism are not related. Maybe, it takes place only for uncharged electromagnetic field, where quantity of particles (photons) isn't equal to quantity of charges, as for electrons.

At rotations (\ref{eq50}), (\ref{eq51}):

\begin{equation}\label{eq88}
\alpha^A_a\sim \varepsilon_{kij}E_j , \;\;\varepsilon_{kij} H_j;\qquad
X^\mu_a=0,
\end{equation}

\noindent so the current is:

\begin{equation}\label{eq89}
M^0_k=l\,S_k=\frac{l}{4\pi}[\vec E\times \vec H]_k,
\end{equation}

\begin{equation}\label{eq90}
M^l_k=\frac{l}{8\pi}\left[\delta^l_k(E^2+H^2)-E^lE_k-H^lH_k\right],
\end{equation}

\noindent and the conservation law is:

\begin{equation}\label{eq91}
\partial_0 S_k+\partial_k\omega -\frac{1}{8\pi}(E^l\partial_l E_k+H^l\partial_l
H_k)=0
\end{equation}

\noindent (when getting it was used that $\partial_k E_k=0$, $\partial_k H_k=0$).

At boosts (\ref{eq50}), (\ref{eq51}):

\begin{equation}\label{eq92}
\alpha^A_a\sim -\varepsilon_{kij}H_j ,  \;\; \varepsilon_{kij} E_j,\qquad X^\mu_a=0,
\end{equation}

\noindent and current is:

\begin{equation}\label{eq93}
L^0_k=0,
\end{equation}

\begin{equation}\label{eq94}
L^l_k=-\frac{l}{8\pi}(E^l H_k-H^l E_k).
\end{equation}

\noindent In such case conservation law leads to identity:

\begin{equation}\label{eq95}
E^l\partial_l E_k=H^l\partial_l H_k
\end{equation}

\noindent (when getting it was also used that $\partial_k E_k=0$,
$\partial_k H_k=0$).\vskip 3mm

\section{Relativistic generalization of $L^D$}

The necessity of relativistic generalization of $L^D$ consists of the following. First, the Lagrangian $L^D$ (\ref{eq46}) has a lower time index 0, that ensures the fact, that it's Rainich current is a zero component of ordinary Maxwell's energy-momentum tensor (\ref{eq76}). Second, not all of Maxwell equations (but only dynamical) are Lagrange equations for $L^D$. And third, it is desirable to find the "reasons" of phase symmetry. It, maybe, would allow to solve the mystery of electron with it's spin, mass, etc.

Firstly, let's write down formulas:

\begin{equation}\label{eq96}
F_{i0}=E_i,\qquad H_i=\frac{1}{2}\varepsilon_{ijk}F^{jk},\qquad
F^{jk}=\varepsilon^{ijk}H_k ,
\end{equation}

\begin{equation}\label{eq97}
\tilde
F_{\mu\nu}=\frac{1}{2}\varepsilon_{\mu\nu\rho\tau}F^{\rho\tau},
\end{equation}

\begin{equation}\label{eq98}
\tilde F_{i0}=H_i,\qquad E_i=-\frac{1}{2}\varepsilon_{ijk}\tilde
F^{jk},\qquad \tilde F^{ij}=-\varepsilon^{ijk}E_k ,
\end{equation}

Lagrangian (\ref{eq46}):

\begin{equation}\label{eq99}
L^D=\frac{l}{8\pi}\left[\vec H\cdot (\partial_0\vec E-\nabla\times
\vec H)-\vec E\cdot(\partial_0\vec H+\nabla\times\vec E)\right].
\end{equation}

\noindent is a zero component of vector Lagrangian:

\begin{equation}\label{eq100}
L^D_\rho =\frac{l}{8\pi}(\tilde F_{\rho\nu}\partial_\mu
F^{\nu\mu}-F_{\rho\nu}\partial_\mu \tilde F^{\nu\mu}).
\end{equation}

\noindent Independent variation of vector Lagrangian $L^D_\rho$ with respect to $F^{\mu\nu}$ and $\tilde F^{\mu\nu}$ lead to pair of equations:

\begin{equation}\label{eq101}
\frac{\delta L^D_\rho}{\delta
F^{\mu\nu}}=-\frac{l}{16\pi}\left[g_{\rho\mu}\partial^\sigma
\tilde F_{\nu\sigma}+\partial_\nu \tilde F_{\rho\mu}-(\mu
<-->\nu)\right]=0,
\end{equation}

\begin{equation}\label{eq102}
\frac{\delta L^D_\rho}{\delta \tilde
F^{\mu\nu}}=\frac{l}{16\pi}\left[g_{\rho\mu}\partial^\sigma
F_{\nu\sigma}+\partial_\nu F_{\rho\mu}-(\mu <-->\nu)\right]=0.
\end{equation}

\noindent Summing by $\rho$ and $\mu$, we get first and second pair of Maxwell equations (\ref{eq43}), (\ref{eq41}):

\begin{equation}\label{eq103}
\partial^\sigma
\tilde F_{\nu\sigma}=0,
\end{equation}

\begin{equation}\label{eq104}
\partial^\sigma
 F_{\nu\sigma}=0,
\end{equation}

\noindent which is the full system of Maxwell equations (besides dynamical equations also equality to zero of divergences $\vec E$ and $\vec H$ ). Are there additional equations in equations (\ref{eq101}) and (\ref{eq102}) in addition to Maxwell equations? In any case the multiplication of (\ref{eq101}) and (\ref{eq102}) by $\varepsilon^{\rho\tau\mu\nu}$ again leads to equations (\ref{eq104}) and (\ref{eq103}) (in reverse order). So each of the equations (\ref{eq101}) or (\ref{eq102}) are enough to get the full system of Maxwell equations (\ref{eq103}) and (\ref{eq104}).

If not to consider $F^{\mu\nu}$ and $\tilde F^{\mu\nu}$ as independent and to use the formula (\ref{eq97}), we'll get:

\begin{equation}\label{eq105}
\frac{\delta L^D_\rho}{\delta
F^{\mu\nu}}=\frac{l}{8\pi}\left[\varepsilon_{\rho\tau\mu\nu}\partial_\sigma
F^{\tau\sigma}-g_{\rho\mu}\partial^\sigma\tilde
F_{\nu\sigma}+g_{\rho\nu}\partial^\sigma\tilde
F_{\mu\sigma}-\partial_\nu\tilde F_{\rho\mu}+\partial_\mu\tilde
F_{\rho\tau}+\varepsilon_{\sigma\tau\mu\nu}\partial^\tau
F^\sigma_\rho \right]=0,
\end{equation}

\noindent in particular, from extra, spatial part of vector Lagrangian $L^D_m$, we have:

\begin{equation}\label{eq106}
\frac{\delta L^D_m}{\delta
F^{ij}}=\frac{l}{8\pi}\left[\varepsilon_{mij}\partial_\sigma
F^{0\sigma}-g_{mi}\partial^\sigma\tilde
F_{j\sigma}+g_{mj}\partial^\sigma\tilde
F_{i\sigma}-\partial_j\tilde F_{mi}+\partial_i\tilde
F_{mj}+\varepsilon_{kij}\partial_k
F_{m0}-\varepsilon_{kij}\partial_0 F_{mk} \right]=0,
\end{equation}

\begin{equation}\label{eq107}
\frac{\delta L^D_m}{\delta
F^{0i}}=\frac{l}{8\pi}\left[\varepsilon_{mik}\partial_\sigma
F^{k\sigma}-g_{mi}\partial_l\tilde F_{l0}-\partial_{i}\tilde
F_{m0}+\partial_0\tilde F_{mi}-\varepsilon_{ikl}\partial^l
F_{mk}\right]=0.
\end{equation}

We know, that variation of $L^D_0$ is enough to get only dynamical part of Maxwell equations. However, variation of spatial part $L^D_m$ is enough to get the full system of Maxwell equations. Really:

\begin{equation}\label{eq108}
\varepsilon_{mij}\frac{\delta L^D_m}{\delta
F^{ij}}=\frac{2\,l}{3\pi}\partial_k E_k=0,
\end{equation}

\begin{equation}\label{eq109}
\delta_{mi}\frac{\delta L^D_m}{\delta
F^{ij}}=-\frac{l}{2\pi}\partial^\sigma \tilde
F_{j\sigma}=\frac{l}{2\pi}(\partial_0
H_j+\varepsilon_{jim}\partial_i E_m)=0,
\end{equation}

\begin{equation}\label{eq110}
\delta_{mi}\frac{\delta L^D_m}{\delta
F^{0i}}=-\frac{3\,l}{4\pi}\partial_k H_k=0,
\end{equation}

\begin{equation}\label{eq111}
\varepsilon_{mij}\frac{\delta L^D_m}{\delta
F^{0i}}=\frac{l}{2\pi}\partial^\sigma
F_{j\sigma}=-\frac{l}{2\pi}(\partial_0 E_{j}
-\varepsilon_{jim}\partial_i H_m)=0.
\end{equation}\vskip 3mm

\section{Conserved currents of vector Lagrangian}

At translations

\begin{equation}\label{eq112}
\delta F_{\mu\nu}=-t^\alpha\partial_\alpha F_{\mu\nu},
\end{equation}

\begin{equation}\label{eq113}
\delta \tilde F_{\mu\nu}=-t^\alpha\partial_\alpha \tilde F_{\mu\nu},
\end{equation}

\begin{equation}\label{eq114}
\alpha^A_a \sim -\partial_\nu F_{\mu\nu}, -\partial_\nu \tilde F_{\mu\nu},\qquad
X^\mu_a \sim \delta^\mu_a,
\end{equation}

\noindent and in consequence of translation invariance of the vector Lagrangian, current

\begin{equation}\label{eq115}
T^\mu_{\rho\alpha}=\frac{l}{8\pi}(\tilde
F_{\rho\tau}\partial_\alpha
F^{\tau\mu}-F_{\rho\tau}\partial_\alpha \tilde
F^{\tau\mu})-\frac{l}{8\pi}\delta^\mu_\alpha (\tilde
F_{\rho\tau}\partial_\sigma F^{\tau\sigma}-
F_{\rho\tau}\partial_\sigma \tilde F^{\tau\sigma}).
\end{equation}
is conserved: $\partial_\mu T^\mu_{\rho\alpha}=0$. On shell the last term disappears, and the conservation equation of "energy-momentum" for $L_\rho ^D$ takes the form:

\begin{equation}\label{eq116}
\partial_\mu \tilde F_{\rho\tau}\partial_\alpha F^{\tau\mu}=\partial_\mu F_{\rho\tau}\partial_\alpha \tilde F^{\tau\mu}.
\end{equation}

Phase transformations:

\begin{equation}\label{eq117}
\delta F_{\mu\nu}=-\tilde F_{\mu\nu}\delta\varphi ,
\end{equation}

\begin{equation}\label{eq118}
\delta \tilde F_{\mu\nu}= F_{\mu\nu}\delta\varphi ,
\end{equation}

\noindent that generalize transformations (\ref{eq16}), lead to the current:

\begin{equation}\label{eq119}
J^\mu_\rho =\frac{l}{8\pi} \left(\tilde F_{\rho\tau}\tilde
F^{\tau\mu}+F_{\rho\tau}F^{\tau\mu}\right)=\frac{l}{4\pi}\left(F_{\rho\tau}F^{\tau\mu}+\frac{1}{4}\delta^\mu_\rho
F_{\alpha\beta}F^{\alpha\beta}\right) ,
\end{equation}

\noindent which coincides (up to the factor $l$) with the energy-momentum tensor (\ref{eq76}) for the Maxwell's Lagrangian.

Let us consider the Lorentz transformation of $ L^D_\rho $. With them:

\begin{equation}\label{eq120}
\delta F_{\mu\nu}=\omega^\sigma_\mu
F_{\sigma\nu}+\omega^\sigma_\nu F_{\mu\sigma},
\end{equation}

\begin{equation}\label{eq121}
\delta \tilde F_{\mu\nu}=\omega^\sigma_\mu \tilde
F_{\sigma\nu}+\omega^\sigma_\nu \tilde F_{\mu\sigma},
\end{equation}

\begin{equation}\label{eq122}
\delta L^D_\rho = \omega^\sigma_\rho L^D_\rho .
\end{equation}

\noindent It follows from (\ref{eq122}) that since on shell $L^D_\rho=0$, we have $\delta L^D_\rho =0$.
\vskip 3mm

\section{Conclusion}

It remains unclear how the Rainich phase $\theta$ relates to the ordinary quantum-mechanical phase of the Hilbert space of quantum electrodynamics.

As to role of a phase in the quantum mechanics, maybe it exists a variant of quantum mechanics, which has only dynamic phase symmetry, as Maxwell's electrodynamics, with the possibility of expression of independent components of $\psi$-function through potentials.

It remains mysterious the existence of vector Lagrangians and the possibility of getting the same physical quantities from different Lagrangians as a result of different symmetries. It would be worth to examine this question in general case.


\end{document}